\documentclass[
aps,
amsmath,amssymb,
prl,
reprint,
groupedaddress,
superscriptaddress,
]{revtex4-2}

\usepackage{graphicx}
\usepackage{dcolumn}
\usepackage{bbm}
\usepackage{bm}
\usepackage{hyperref}
\hypersetup{colorlinks=true,citecolor=blue,urlcolor=blue,linkcolor=blue}

\begin{document}


\title{Defocus-integration interferometric scattering microscopy for speckle suppression and enhancing nanoparticle detection on substrate}

\author{Nanfang Jiao}
\altaffiliation{These authors contributed equally to this work.}
\affiliation{School of Physics and Wuhan National Laboratory for Optoelectronics, Huazhong University of Science and Technology, Luoyu Road 1037, Wuhan, 430074, People's Republic of China}

\author{Shupei Lin}
\altaffiliation{These authors contributed equally to this work.}
\email[To whom correspondence should be addressed:\\ ]{shupei\_lin@hust.edu.cn}
\affiliation{School of Physics and Wuhan National Laboratory for Optoelectronics, Huazhong University of Science and Technology, Luoyu Road 1037, Wuhan, 430074, People's Republic of China}


\author{Delong Feng}
\affiliation{School of Physics and Wuhan National Laboratory for Optoelectronics, Huazhong University of Science and Technology, Luoyu Road 1037, Wuhan, 430074, People's Republic of China}

\author{Yong He}
\affiliation{School of Physics and Wuhan National Laboratory for Optoelectronics, Huazhong University of Science and Technology, Luoyu Road 1037, Wuhan, 430074, People's Republic of China}

\author{Xue-Wen Chen}
\email[To whom correspondence should be addressed:\\ ]{xuewen\_chen@hust.edu.cn}
\affiliation{School of Physics and Wuhan National Laboratory for Optoelectronics, Huazhong University of Science and Technology, Luoyu Road 1037, Wuhan, 430074, People's Republic of China}
\affiliation{Institute for quantum science and engineering, Huazhong University of Science and Technology, Luoyu Road 1037, Wuhan, 430074, People's Republic of China}

\begin{abstract}
Direct optical detection and imaging of single nanoparticles on substrate in wide field underpin vast applications across different research fields. 
However, the speckles originating from the unavoidable random surface undulations of the substrate ultimately limit the size of the decipherable nanoparticles by the current optical techniques, including the ultrasensitive interferometric scattering microscopy (iSCAT). Here we report a defocus-integration iSCAT to suppress the speckle noise and to enhance the detection and imaging of single nanoparticles on ultra-flat glass substrate and silicon wafer. In particular, we discover distinct symmetry properties of the scattering phase between the nanoparticle and the surface undulations that cause the speckles. Consequently, we develop the defocus-integration technique to suppress the speckles.We experimentally achieve an enhancement of the signal to noise ratio by 6.9 dB for the nanoparticle detection. We demonstrate that the technique is generally applicable for nanoparticles of various materials and for both low and high refractive-index substrates.
\end{abstract}

\maketitle

Direct optical detection and imaging of individual nanoparticles in wide field are important for many applications in a variety of research fields ranging from life science to semiconductor industry \cite{Kukura2021ChemRev}. For nanoparticles in dynamic motion, a number of direct optical microscopy techniques can provide exceptionally good sensitivity thanks to the applicability of subsequent background subtraction which could successfully remove the static background \cite{Matthewnnano,Zhang2020NM,Dey2023acsphch}. For instance, the interferometric scattering  microscopy (iSCAT) is capable of directly detecting and quantifying individual unlabeled proteins and the other individual biological molecules in solution \cite{Marek2014NC,6hsieh2014tracking,4young2018quantitative,Milan2021SmallMethod,5Qiang2022PNAS}. However, the detection of tiny stationary nanoparticles adhered to a substrate remains a challenge because the method of subsequent background subtraction becomes inapplicable and the signal from the nanoparticle could be easily masked by the speckle noises. Indeed, it has been reported that even ultra-flat substrates with sub-nanometer surface undulations, such as polished glass coverslips \cite{7chada2015glass} or silicon wafers \cite{8malik1993surface}, can generate speckles that hinders the detection of nanoparticles with diameters of tens of nanometers depending on the level of surface roughness \cite{9van2011particle,background2017ACS,12lin2022optical}. In this letter, we demonstrate a defocus-integration iSCAT  technique to suppress the speckle noise and to enhance the sensitivity of detecting nanoparticles adhered to an ultra-flat dielectric substrate.

\begin{figure}
	\centering
	\includegraphics[width=8.6cm]{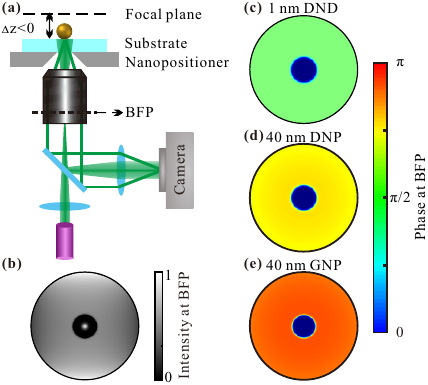}
	\caption{(a) Schematic of the experimental setup. The defocus ($\Delta z$) is scannable. (b) Calculated light intensity distribution at the BFP for the DND (diameter 100 nm, height 1 nm) with $\Delta z$ = 0 on a perfectly flat glass coverslip surface. The numerical aperture of the objective is 0.9. The light intensity within the black circle is multiplied by $10^{-8}$. The calculated phase distributions of the electric field at the BFP for (c) a 1 nm DND, (d) a 40 nm DNP and (e) a 40 nm GNP.
	}
	\label{FIG1}
\end{figure}

We start the discussion with a schematic diagram of an iSCAT setup illustrated in  Fig. \ref{FIG1}(a). The iSCAT contrast originates from the interference between the reflected reference beam from the substrate and the scattered light from the sample as
\begin{equation}
c = \frac{|\textbf{E}_{ref}+\textbf{E}_{sca}|^2}{|\textbf{E}_{ref}|^2}-1
      \approx\frac{2E_{sca}\cos\Delta\phi}{E_{ref}}
\label{eq:1}
\end{equation}
where $E_{ref}$ and $E_{sca}$ represent the electric field amplitudes of reference and scattering, respectively. $\Delta\phi=\phi _{sca}-\phi _{ref}$ denotes the phase difference between the two fields. $\Delta\phi$ depends on various factors, including the refractive indices of the scatterer and the substrate, the distances between the center of the scatterer and the substrate surface, as well as the defocus position $\Delta\textit{z}$ \cite{13he2021multiscale}. The symmetry property of phase change over the defocus in iSCAT is the key to speckle suppression. In the following, we separately examine this property for the scatterings from the tiny surface undulations that cause the speckles and the single nanoparticles we aim to detect. 

An ultra-flat substrate typically exhibits sub-nanometer surface undulations that laterally could extend up to hundreds of nanometers, forming domains of tiny hills and valleys \cite{12lin2022optical}. Therefore, a typical domain can be effectively approximated by a thin dielectric nano-disk (DND) with a diameter of about one hundred nanometers and a height of one nanometer. This approximation helps to gain  physical insight of the scattering from the sub-nanometer surface undulations of the substrate.

By applying the multiscale iSCAT model \cite{13he2021multiscale}, we could track both the intensity and phase of the fields along the whole optical path, in particular, including the Fourier plane (corresponding to the back focal plane (BFP) of the objective in experiment). For the calculations, we assume the normal illumination is a Gaussian beam polarized along {\it x}-direction with a beam waist of 3 $\mu$m with a wavelength of 520 nm. Fig. \ref{FIG1}(b) displays the calculated light intensity distribution at the Fourier plane for a DND on a perfectly flat coverslip. The bright spot at the center represents the contribution by the reflected reference beam.The part outside the circle is due to the scattering, which essentially corresponds to the radiation of an in-plane electric dipole above an interface \cite{16hecht2012principles}. Here we assume the use of an air objective with a numerical aperture (NA) of 0.9, which defines the size of the circle. Additional calculations for  gold nanoparticles (GNPs) (40 nm), and dielectric nanoparticles (DNPs) (40 nm) show essentially the identical intensity patterns at the Fourier plane as in Fig. \ref{FIG1}(b).

\begin{figure}[t]
	\centering
	\includegraphics[width=8.6cm]{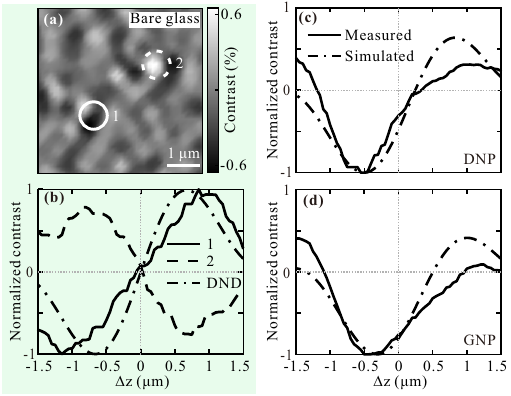}
	\caption{(a) Measured interference contrast image of a bare glass coverslip at $\Delta z$ = -800 nm. (b) The solid and dashed lines are the measured axial evolution of the normalized contrast at the two spots selected in (a); The dot-dashed line is the calculated curve for a DND on a perfectly-flat coverslip; (c) Measured (solid line) for the case of a 60 nm DNP and numerical calculated (dot-dashed line) for the case of a 40 nm DNP. (d) Measured (solid line) and simulated (dot-dashed line) for the case of a 40 nm GNP. 
	}
	\label{FIG2}
\end{figure}

Fig. \ref{FIG1}(c) $\sim$ \ref{FIG1}(e) demonstrate the calculated phase distributions of the electric field at BFP when $\Delta z = 0$ for the cases of the  DND,  DNP and GNP, respectively. The refractive indices of the DND and DNP are set to 1.52 and 1.60, which match that of glass and polystyrene, respectively. For gold, a complex refractive index of $n_{\rm{Au}}$  = 0.63+ 2.07i is used for the illumination wavelength of 520 nm \cite{15johnson1972optical}. The averaged phase difference between the scatterings from the DND, DNP and GNP, and the reference beam, are found to be about $\pi$/2, 0.65$\pi$, and 0.76$\pi$, respectively. For the DNP, in comparison to the DND, an additional phase shift of $2k_0r = 0.15\pi$ is introduced, where $k_0$ is the wavenumber of the illumination light in vacuum, $r$ = 20 nm is radius of the DNP and is also the distance from the center of the DNP to the coverslip surface. For the 40 nm GNP, besides a phase shift of 0.15$\pi$ induced by the size, there is another phase shift of 0.11$\pi$ caused by the complex dielectric constant of gold. As we show below, the initial phase difference at $\Delta z = 0$ is an important parameter. When the substrate is positioned at a particular defocus distance $\Delta z$ other than 0, each planewave component at the Fourier plane (i.e., BFP) will acquire a phase \cite{Avci2016OE}
\begin{equation}
\phi(x_b,y_b,\Delta z)=nk_0\Delta z\cos\theta 
\label{eq:2}
\end{equation}
where $n$ is the refractive index of medium between the objective and the substrate. One has  $\Delta z<0$ when the sample leaves from the focal plane towards the objective, as shown in Fig. \ref{FIG1}(a). The angle $\theta$ is a quantity associated with the position $(x_b,y_b)$ at BFP according to $\sin\theta=\sqrt{{x_b}^2+{y_b}^2}/f_o$, where $f_o$ is the effective focal length of the objective. Since the scattered light and reference beam possess distinct intensity distributions at BFP, a defocus $\Delta z$ would evidently cause an additional phase difference $\Delta\phi(\Delta z)=\phi_{in}+\bar{\phi}_{s}-\phi_{r}$, where $\phi_{in}$ is the initial phase difference at $\Delta z=0$,  $\bar{\phi}_{s}$ is the phase in Eq. (\ref{eq:2}) averaged over the BFP for the scattered field and $\phi_{r}=\phi(0,0,\Delta z)$ is the phase for the reference beam.

For the case of DND with $\phi_{in}=\pi/2$, we deduce $\Delta\phi(-\Delta z)=\pi-\Delta\phi(\Delta z)$ for an arbitrary $\Delta z$. As a result, based on Eq. (\ref{eq:1}), the interference contrast of the 1 nm DND should satisfy $c_{DND} (\Delta z) = -c_{DND} (-\Delta z)$. It implies that the axial evolution of the interference contrast as the defocus for the 1 nm DND should show anti-symmetry. Therefore, the axial evolution of speckles should also exhibit anti-symmetry, as the sub-nanometer undulation of the substrates share the same physics as the DND. However, for the case of nanoparticles, either dielectric or metallic, we have $\phi_{in}> \pi/2$ and thus have $\Delta\phi(-\Delta z)\neq\pi-\Delta\phi(\Delta z)$. Consequently, the axial evolution of the interference contrasts for nanoparticles does not possess the anti-symmetry as the DND. Moreover, as $\phi_{in}$ becomes closer to  $\pi$, the defocus value corresponding to the contrast dip will be closer to $\Delta z=0$ compared to the defocus corresponding to the contrast peak. In addition, the amplitude of the contrast dip will be larger than that of the peak. These factors make the contrast evolution over the defocus deviate more from anti-symmetry.

We next show the comparison between the experimental measurements and numerical calculations to validate the theoretical predictions. Our experimental setup is a common path iSCAT microscope, as illustrated in Fig. \ref{FIG1}(a). A laser beam with a central wavelength of 520 nm and a bandwidth of 40 nm is focused onto the BFP of the objective, thereby generating normal incidence wide-field illumination on the sample. The objective collects both the reference beam and the scattered light. The collected lights subsequently pass through a beam splitter and a tube lens to form an interference contrast image on the camera. The sample is positioned on a three-dimensional nanopositioner, enabling the selection of defocus positions and measurement areas. In experiment, we firstly determine the position of the focal plane by comparing the measured axial evolution of interference contrasts with the calculated ones for large size nanoparticles. Then the position is marked by the distance from the focal plane to the defocus position where a smallest confocal spot is obtained without the wide field lens in the optical path.

Fig. \ref{FIG2}(a) presents the measured interference contrast image for an area of a bare glass coverslip at $\Delta z$ = -800 nm. Two characteristic points (with interference contrasts significantly greater than 0 and significantly less than 0, respectively) are selected to plot the axial evolution of their interference contrasts in Fig. \ref{FIG2}(b). The dot-dashed line is the calculated curve for the 1 nm DND on a perfectly-flat glass coverslip. We can see that the peak values of the three curves are nearly equal to the absolute values of their corresponding dips, and the distances from the peaks and dips to the focal plane are also nearly equal. All three curves exhibit excellent anti-symmetry. An anti-symmetry coefficient for an axial evolution curve can be defined as
\begin{equation}
\textit{F} = 1-\left|\frac{\int_{-z_0}^{z_0}cdz}{\int_{-z_0}^{z_0}|c|dz}\right|
\label{eq:3}
\end{equation}
Here, $\textit{F} = 1$ indicates complete anti-symmetry. The anti-symmetry coefficients of the three curves are 0.95, 0.97, and 0.99, respectively, demonstrating good agreement with the theoretical analysis.

In Fig. \ref{FIG2}(c), the dot-dashed line represents the calculated axial evolution curve of the interference contrast for the 40 nm DNP, while the solid line is the measured result for a 60 nm polystyrene nanoparticle. The choice of 60 nm in the experiment, instead of the 40 nm DNP used in simulation, is due to practical consideration. A 60 nm DNP yields approximately 3.3 times the interference contrast compared to a 40 nm DNP, ensuring a higher signal-to-noise ratio (SNR) in the measurement to prevent the curve from being distorted by speckle. Fig. \ref{FIG2}(c) shows that the dip and peak values of the curves occur at $\Delta z$ = -500 nm and $\Delta z$ = 750 nm, respectively, both in measurement and numerical simulation. Further more, the peak values of the curves are only 25\% and 64\% of the corresponding absolute dip values in measurement and simulation, respectively. The anti-symmetry coefficients of the axial evolution of the interference contrasts of the DNPs are 0.69 and 0.64 in measurement and in simulation, respectively, much weaker than that of the speckle contrasts.

\begin{figure}[t]
	\centering
	\includegraphics[width=8.6cm]{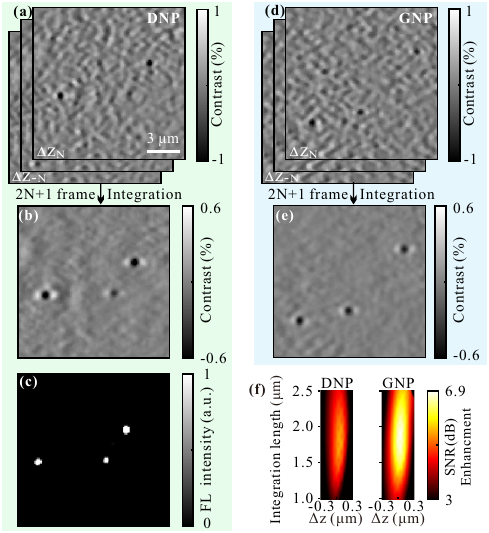}
	\caption{Interference contrast images of (a) iSCAT at one particular position of defous and (b) our defocus-integration iSCAT for a glass coverslip surface with 40 nm DNPs dopped with fluorescent molecules. (c) Fluorescence (FL) image of the same area of the coverslip. Interference contrast images obtained with (d) usual iSCAT and (e) our DI-iSCAT for the glass coverslip with 30 nm GNPs. (f) SNR enhancement maps as functions of defocus center and integration length for the DNPs and GNPs.
	}
	\label{FIG3}
\end{figure}

Obviously, for the GNP with a larger initial phase, we anticipate that the evolution curve of the contrast will deviate more from anti-asymmetry. In Fig. \ref{FIG2}(d), both the measurement and simulation reveal that the dip and peak positions of the interference contrast occur at $\Delta z$ = -300 nm and $\Delta z$ = 950 nm, respectively. The peak values of the interference contrasts for the GNPs are only 9\% and 41\% of the absolute dip values of that in measurement and simulation, respectively. The anti-symmetry coefficients of the measured and simulated curves for the GNPs are 0.24 and 0.45, respectively, smaller than that for the DNPs of the same sizes. Once again, the experimental result agrees well with the theoretical analysis.

\begin{figure}
	\centering
	\includegraphics[width=8.6cm]{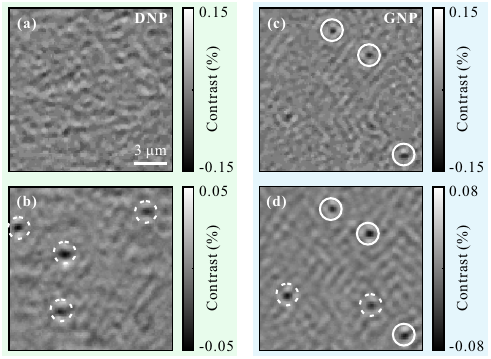}
	\caption{Interference contrast images obtained from (a) usual iSCAT and (b) our DI-iSCAT for 40 nm DNPs on a silicon wafer surface. Interference contrast images obtained from (c) usual iSCAT and (d) our DI-iSCAT for 30 nm GNPs on the silicon wafer surface.
	}
	\label{FIG4}
\end{figure}

The axial evolutions of the speckle contrasts exhibit much better anti-symmetry than that of the nanoparticles. Consequently, we can suppress speckle by taking an integration on a series of measured defocus scanning interference contrast images from a negative defocus position to the corresponding positive defocus position, respectively, as shown in Fig. \ref{FIG3}. The integration process can be represented as:
\begin{equation}
\bar{c}=\frac1{2N+1}\sum_{i=-N}^{N}c(i\delta z)
\label{eq:4}
\end{equation}
where $\delta z$ is the step size of defocus scanning in experiment, and $1/(2N+1)$ is the normalized factor for the defocus integration length of $2N\delta z$.

Fig. \ref{FIG3}(a) and \ref{FIG3}(d) display the measured raw interference contrast images for the 40 nm DNPs, which are doped with fluorescent molecules, and the 30 nm GNPs on glass coverslips, respectively. Defocus scanning was performed with a step size of 100 nm over a range of 6 $\mu$m. It is challenging to identify the nanoparticles in both Fig. \ref{FIG3}(a) and \ref{FIG3}(d). Fig. \ref{FIG3}(b) and \ref{FIG3}(e) represent the defocus-integration iSCAT (DI-iSCAT) contrast images obtained by applying Eq. (\ref{eq:4}) to Fig. \ref{FIG3}(a) and \ref{FIG3}(d), respectively, with a defocus integration length of 1.9 $\mu$m. In both figures, three particle-like point spread functions can be clearly observed. Fig. \ref{FIG3}(c) is the fluorescence image for the same area as in Fig. \ref{FIG3}(a). By the comparison between Fig. \ref{FIG3}(b) and \ref{FIG3}(c), we can confirm that the distinct particle-like point spread functions in Fig. \ref{FIG3}(b) and \ref{FIG3}(d) are from targeted nanoparticles.

We define the SNR of detecting nanoparticle as: $\rm{SNR}=10\log_{10}c_s/c_n$, where $c_s$ represents the absolute value of the interference contrast of the nanoparticle, and $c_n$ represents the standard deviation of the speckle contrasts. The comparison between the raw iSCAT contrast images and the DI-iSCAT contrast images reveals that the defocus integration process enhances the SNRs by 5.4 dB for DNPs and 6.9 dB for GNPs, respectively. The greater SNR enhancement for GNPs should be attributed to the weaker anti-symmetry in the axial evolution of the interference contrasts. Fig. \ref{FIG3}(f) shows the SNR enhancement maps for the DNPs and GNPs, which are functions of defocus integration lengths and central positions of defocus. We can observe that the SNR enhancements significantly decrease when the center of the defocus range deviates from $\Delta z=0$ because the centers of anti-symmetry of speckle interference contrasts locate at $\Delta z=0$. On the other hand, the SNR enhancements are relatively robust to change in the length of the defocus integration, as it does not significantly alter the anti-symmetry of the interference contrasts.

Silicon wafer is the most important substrate in semiconductor industry. We also perform the speckle suppression for detecting nanoparticles on silicon wafers. Fig. \ref{FIG4}(a) and \ref{FIG4}(b) are the measured raw iSCAT contrast image and DI-iSCAT contrast image a silicon wafer with 40 nm DNPs, respectively. There is only the presence of speckle noise shown in Fig. \ref{FIG4}(a). Four DNPs (indicated by dashed circles), that are indecipherable in Fig. \ref{FIG4}(a), are clearly displayed in Fig. \ref{FIG4}(b). These nanoparticles are further confirmed by the corresponding fluorescence image of the same area (not shown here).  Similarly, the measured raw iSCAT contrast image and DI-iSCAT contrast image for a silicon wafer with 30 nm GNPs are shown in Fig. \ref{FIG4}(c) and \ref{FIG4}(d), respectively. In Fig. \ref{FIG4}(c), only three GNPs (indicated by solid circles) can be recognized with very low SNRs. However, in Fig. \ref{FIG4}(d), these three GNPs are displayed much more clearly, and two additional GNPs (indicated by dashed circles) can be easily identified.

In summary, we have demonstrated a defocus-integration interferometric scattering microscopy to suppress the inevitable speckle noise originating from the subnanometer surface undulations of the substrate and therefore enhanced the detection and imaging of single nanoparticles adhered to substrate. We identify from theoretical analysis that the evolution of the speckle contrasts over the defocus possesses the property of anti-symmetry and therefore an integration over the defocus could suppress the speckle. Different from the speckles, the signal from the nanoparticles does not possess such anti-symmetry. Consequently, an integration operation will make the signal emerge from the speckle noises. Our experiments successfully confirm the theoretical analysis and achieve an enhancement of the detection sensitivity by 6.9 dB. The technique is simple without the need of any hardware change and is applicable to different kinds of nanoparticles and substrate materials. The detection sensitivity may be further improved with the correction of the aberration of the imaging system using adaptive optics techniques \cite{Booth2007PTRSA}. We anticipate the defocus-integration interferometric scattering microscopy technique finds applications in life science and semiconductor industry.

We acknowledge financial support from the Postdoctoral Fellowship Program
of CPSF (Grant Number GZC20230887, S.L.) and National Natural Science Foundation of China (Grant No. 92150111, 62235006, X.-W. C) and the Science and Technology Department of Hubei Province, China (Project No. 2022BAA018, X.-W.C.) and Huazhong University of Science and Technology.

\bibliographystyle{apsrev4-2}
\bibliography{ref.bib}

\end{document}